\documentclass[a4paper,10pt]{article}
\usepackage{amsfonts}
\usepackage{amsmath}
\usepackage{amssymb}
\usepackage[dvips]{graphicx}

\DeclareGraphicsExtensions{.bmp,.png,.pdf,.jpg}

\begin{document}

\title{An estimation of the Moon radius by counting craters: a
generalization of Monte-Carlo calculation of $\pi $ to spherical geometry }
\author{J. S. Ardenghi$^{\dag }$\thanks{%
email:\ jsardenghi@gmail.com, fax number:\ +54-291-4595142} \\
%EndAName
$^{\dag }$IFISUR, Departamento de F\'{\i}sica (UNS-CONICET)\\
Avenida Alem 1253, Bah\'{\i}a Blanca, Buenos Aires, Argentina}
\maketitle

\begin{abstract}
By applying Monte-Carlo method, the Moon radius is obtained by counting
craters in a spherical square over the surface of it. As it is well known,
approximate values for $\pi $ can be obtained by counting random numbers in
a square and in a quarter of circle inscribed in it in Euclidean geometry.
This procedure can be extend it to spherical geometry, where new relations
between the areas of a spherical square and the quarter of circle inscribed
in it are obtained. When the radius of the sphere is larger than the radius
of the quarter of circle, Euclidean geometry is recovered and the ratio of
the areas tends to $\pi $. Using these results, theoretical deviations of $%
\pi $ due to the Moon radius $R$ are computed. In order to obtain this
deviation, a spherical square is selected located in a great circle of the
Moon. The random points over the spherical square are given by a specific
zone of the Moon where craters are distributed almost randomly. Computing
the ratio of the areas, the deviation of $\pi $ allows us to obtain the Moon
radius with an intrinsic error given by the finite number of random craters.
\end{abstract}

\section{Introduction}

The Moon is always a subject of intense debate, for instance, it is the
cause of many natural phenomena, the most common of which are solar eclipses
and ocean tides. In turn, it is a source for testing theories of gravitation
and to investigate geophysical phenomenas (\cite{ha} and \cite{cook}).
Moreover, there are several relational properties between the Moon and the
Earth, such as the instantaneous distance between them or the relative
rotation, as well as intrinsic properties of the Moon, such as the Moon
diameter, its excentricity or the more perplexing long-standing puzzle of
its origin, that requires sophisticated and non-sophisticated experimental
procedures in order to obtain approximate values or conceptual explanations
(see for instance \cite{pell}, \cite{lo}, \cite{linc},\cite{steve}, \cite%
{gor} and \cite{jay}). In turn, the Moon is suitable to desing simple
experimental procedures to obtain accurate information of different
parameters of it (\cite{nasa}), or for example by measuring the Moon's orbit
by using a hand-held camera (see \cite{benja}). In the same line of thought,
this work introduces a novel procedure to estimate the Moon's radius without
using any sophisticated measuring device. Briefly, the method consists in
using the Monte-Carlo method to obtain $\pi $ using random points (see \cite%
{don} and \cite{bloch}) generalized to spherical geometry.\footnote{%
In \cite{bloch}, page 302 there is a plot of $\pi $ in a sphere, where the
value is computed with the ratio of the cimcurference of a circle to its
diameter, measured on the curved surface of the sphere.} In this manuscript,
the sphere will be the Moon and the random points will be the craters on it.
The quarter of circle inscribed in a square on the plane, where the random
points are located and which is used in the typical Monte-Carlo method, is
generalized to a quarter of circle inscribed in a spherical square over the
surface of a sphere. This mathematical generalization of the Monte-Carlo
method to spherical geometry is very simple and implies a derivation of
areas of circles and squares drawn over a spherical surface. Should be
expected that when the radius of the spherical surface tends to infinity,
the Monte-Carlo method gives $\pi $ because spherical geometry in this limit
tends to flat geometry. This leads to the following conclusion: by counting
random points over quarter of circles and squares in spherical geometry we
obtain deviations from $\pi $ by computing the areas ratio and these
deviations are a function of the radius of the sphere. Interestingly, we can
apply the Monte-Carlo method over a quarter of circle in the spherical
surface without knowing the sphere radius but it can be obtained by counting
random points on it. Of course, to do so we must know the function that
relates the deviation from $\pi $ with the spherical surface radius. Using
the craters of the Moon as a random points in a particular spherical square
inscribed on it, we can obtain the deviation from $\pi $ and by using the
theoretical function that relates the deviation from $\pi $ with the Moon's
radius, we can deduce the Moon's radius with an accuracy related to the
number of craters considered. Should be stressed that altough the conceptual
procedure is simple, spherical geometry introduce subtetlies concerned with
circle and squares inscribed over the surface of a sphere that must be taken
into account in order to obtain the correct limit when the flat geometry is
recovered, which can be implemented easily by taking the infinite limit of
the sphere radius. Then, in order to be consise and self-contained, this
mansuscript will be organized as follows:\ In Section II, the function that
relates the deviation from $\pi $ in the Monte-Carlo method applied to
spherical geometry and the spherical surface radius is obtained. In Section
III we used the results obtained in Section II to determine the Moon's
radius with the respective error by considering random craters in a
particular zone of the Moon surface. In last section, the conclusions are
presented and in Appendix some mathematical details are shown.

\section{Monte-Carlo method in spherical geometry}

The Monte-Carlo method is a general method that allows us to obtain specific
results by manipulating random variables (\cite{bro}, \cite{shi}, \cite{duf}%
, \cite{pi1}, \cite{eric}, \cite{ok} and \cite{moh}). As it is well known,
approximate values of $\pi $ can be obtained by counting random numbers on a
square and on a quarter of circle inscribed in it in a plane(\cite{wil}). 
\begin{figure}[tbh]
\begin{minipage}{0.47\linewidth}
\includegraphics[width=80mm,height=59mm]{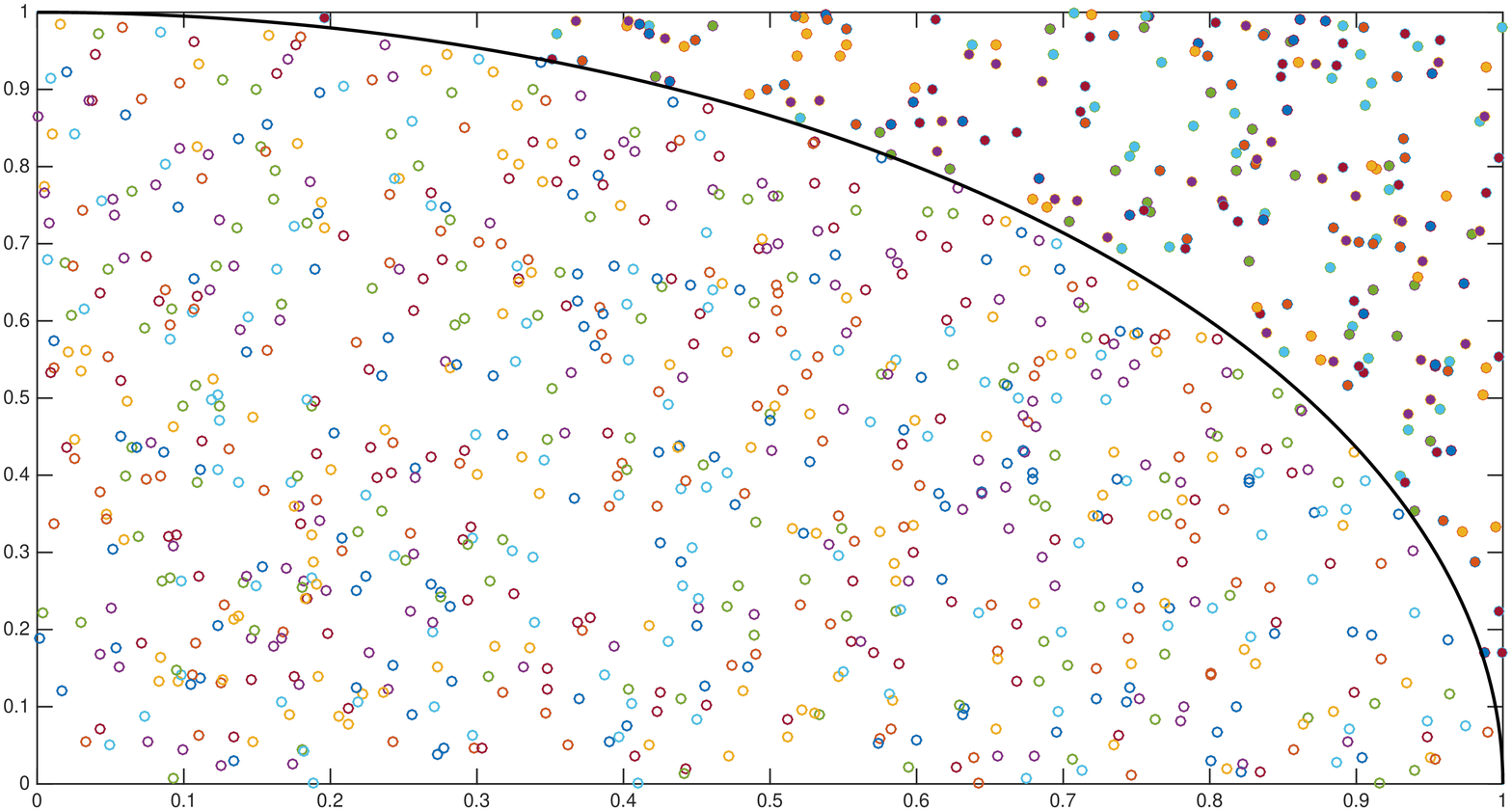} 
%\caption{Random vectors in the unit square.}

\end{minipage}%\quad 
\hspace{0.07cm} 
\begin{minipage}{0.55\linewidth}
\centering
\includegraphics[width=80mm,height=57mm]{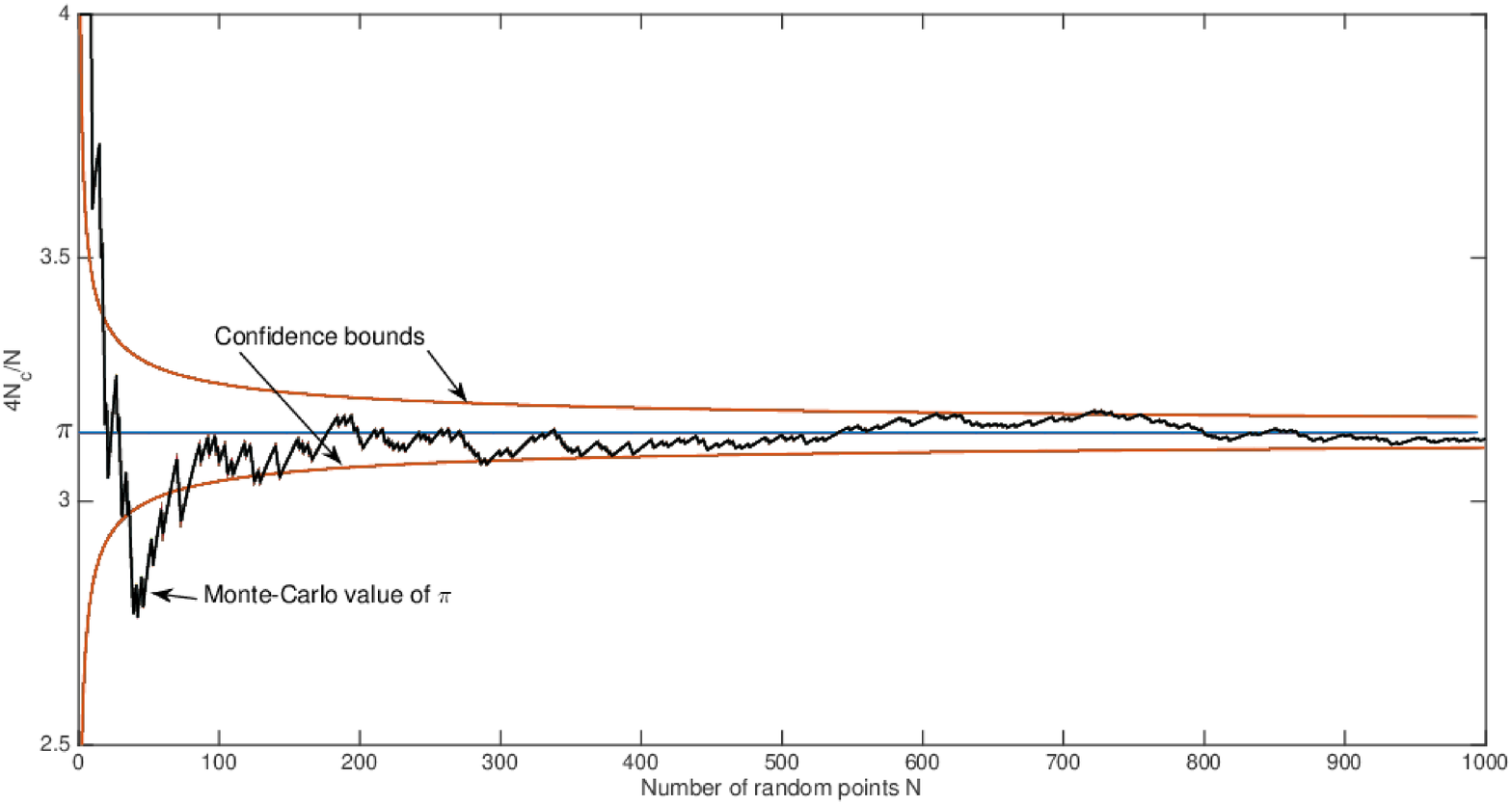}
%\caption{Value of Pi obtained as a function of $N$ random vectors in the unit square.}

\end{minipage}%\label{pdosA} \label{pdos}
%\label{pro}
\caption{Left: Random points in the unit square. Inside the unit square, the
inscribed quarter of circle can be seen. Right: Value of Pi obtained as a
function of $N$ random points. The confidence bounds that behaves as the
square roots of $N$ is shown. }
\label{random}
\end{figure}
In this method, by considering a square of side $L=r$ and a a quarter of
circle of radius $r$ with center in one vertex of the square as it can be
seen in figure \ref{random}, the ratio between the areas read%
\begin{equation}
\frac{A_{C}}{A_{S}}=\frac{\pi r^{2}}{4r^{2}}=\frac{\pi }{4}  \label{1}
\end{equation}%
Considering $N~$random points in the unit square $\left[ 0,1\right] \times %
\left[ 0,1\right] $, the area of this square can be approximated by $N$ and
the area of the quarter-circle can be approximated by the number of random
points $N_{C}$ that lie inside it. By computing $4N_{C}/N$, an approximate
value of $\pi $ can be obtained and the approximation can be accurate by
increasing $N$ as it can be seen in figure \ref{random}, where in turn the
confidence bounds are shown that scales as $N^{-1/2}$ for large $N$. 
\begin{figure}[tbp]
\centering\includegraphics[width=80mm,height=70mm]{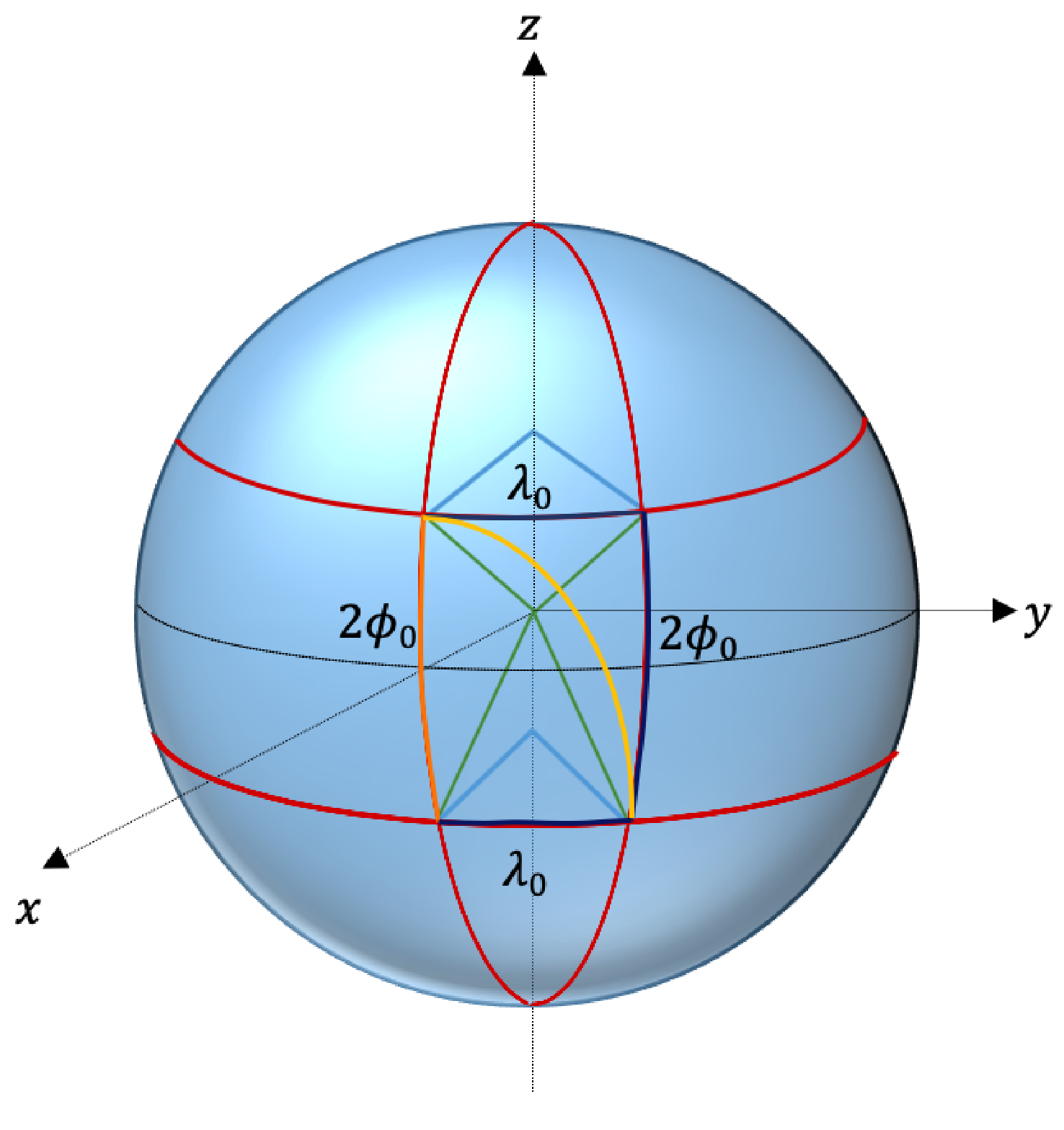}
\caption{Spherical square (blue line) located symmetrically with respect a
great circle of the sphere and the quarter of circle inscribed in it (yellow
line). The sides of the spherical square are shown as $\protect\lambda _{0}$
and $2\protect\phi _{0}$. As it was explained in the introduction, these two
sides must be identical in order to have a correct limit in flat geometry. }
\label{sphere}
\end{figure}
The same argument can be applied in spherical geometry, where we can
consider a square of side $r$ inscribed in a sphere of radius $R$ and a
circle of radius $r$ inscribed in the square as it can be seen in the figure %
\ref{sphere}, where the spherical square is located symmetrically with
respect a great circle of the sphere. This choice of the spherical square
location is suitable when the limit of infinite radius is taken, because the
ratio of areas tends to $\pi $ and the radius of the circle measured over
the surface of the sphere is identical to the length of the sides of the
square.

In \ order to obtain the square and circle areas, we can consider the
surface element area $dA=R\sin \theta d\theta d\varphi $ in spherical
coordinates, where $R$ is the sphere radius, $\varphi $ is the azimuth and $%
\theta $ is the angle with respect the $z$ axis. By using spherical
coordinates it is possible to show that the area of the square of radius $r$
in a sphere of radius $R$ reads (see Appendix)

\begin{figure}[tbp]
\centering\includegraphics[width=100mm,height=70mm]{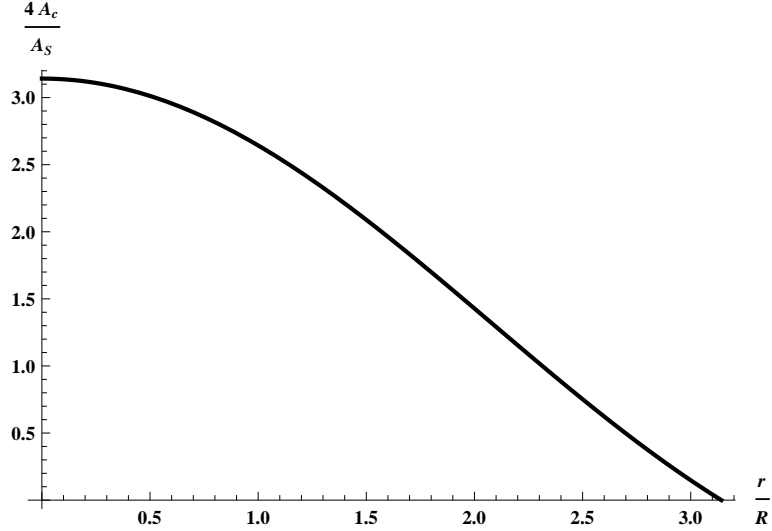}
\caption{Monte-Carlo method to obtain deviations of $\protect\pi $ as a
function of $\frac{r}{R}$ in spherical geometry, where $r$ is the radius of
the circle and the side of the spherical square and $R$ is the radius of the
sphere.}
\label{alfa1}
\end{figure}
\begin{figure}[tbp]
\centering\includegraphics[width=80mm,height=70mm]{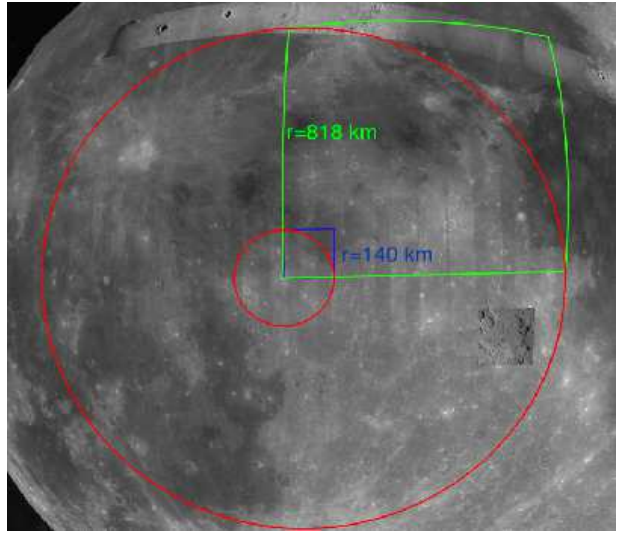}
\caption{Different zones over a spherical surface where the Monte-Carlo
method can be applied. As it can be seen, for $r=140$ km the Monte-Carlo
approximates accurately to flat geometry when the ratio of areas are
computed, where the result expected is $0.998\protect\pi $. The larger zone
with $r=818$ km deviates from flat geometry and the Monte-Carlo method gives 
$0.964{\protect\pi}$.}
\label{moonf}
\end{figure}

\begin{equation}
A_{S}=2rR\tan (\frac{r}{2R})  \label{2}
\end{equation}%
and the area of a circle of radius $r$ inscribed in a sphere of radius $R$
reads%
\begin{equation}
A_{C}=\frac{\pi }{2}R^{2}\left[ 1-\cos (\frac{r}{R})\right]  \label{3}
\end{equation}%
when $r/R\rightarrow 0$ both areas are $A_{S}=r^{2}$ and $A_{C}=\frac{\pi
r^{2}}{4}$ as it is expected and eq.(\ref{1}) holds.\ This limit corresponds
to the case in which $R>r$ (the radius of the sphere is much larger than the
radius of the circle inscribed in the sphere) and the Euclidean geometry is
an accurate approximation. Then, by using eq.(\ref{2}) and eq.(\ref{3}), the
factor $4A_{C}/A_{S}$ reads%
\begin{equation}
\frac{4A_{C}}{A_{S}}=\frac{\pi R}{r}\sin (\frac{r}{R})  \label{3.1}
\end{equation}%
and the result depends on the ratio $r/R$ as it can be seen in the figure %
\ref{alfa1}, where $4A_{C}/A_{S}$ is plotted against $\frac{r}{R}$, where
the limit $r/R\rightarrow 0$ can be seen. Around this limit, the function
behaves as%
\begin{equation}
\frac{4A_{C}}{A_{S}}\sim \pi -\frac{2}{3}\pi (\frac{r}{2R})^{2}+\frac{2}{15}%
\pi (\frac{r}{2R})^{4}+O\left( (\frac{r}{2R})^{6}\right)  \label{3.2}
\end{equation}%
The last equation shows the deviations from $\pi $ in the Monte-Carlo method
applied in spherical geometry and this deviation can be expanded in powers
of $\frac{r}{2R}$, which implies that by knowing how much we have deviate
from $\pi $ then we can deduce $R$ by an appropiate choice of $r$. We will
apply the results obtained in this section in a particular case where nature
provide us with natural random points in a spherical surface: the Moon and
its craters. For instance, in order to see the physical implication of last
equation, in figure \ref{moonf}, two different zones where the Monte-Carlo
method can be implemented in a spherical surface.

In this figure, we are showing the Moon's surface, but any spherical surface
is valid for the discussion. The two different zones are given with the
respective $r$ radius. It is instructive to note that for the smallest $r$
and considering that $R=1737$ km for the Moon's radius, $r/R\sim 0.08$ and
for the second zone with largest $r$, $r/R$ $\sim 0.47$. The first zone with
smallest $r$ implies a small deviation from $\pi $ than the second zone when
the ratios of areas is computed. In the small zone, the Monte-Carlo method
is indistinguishable from the Monte-Carlo in flat geometry, $\frac{4A_{C}}{%
A_{S}}\sim \pi -\frac{2}{3}\pi (0.04)^{2}\sim 0.998\pi $ and in the second
zone $\frac{4A_{C}}{A_{S}}\sim 0.964\pi $.

\section{Estimation of the Moon's radius}

\begin{figure}[tbp]
\centering\includegraphics[width=80mm,height=80mm]{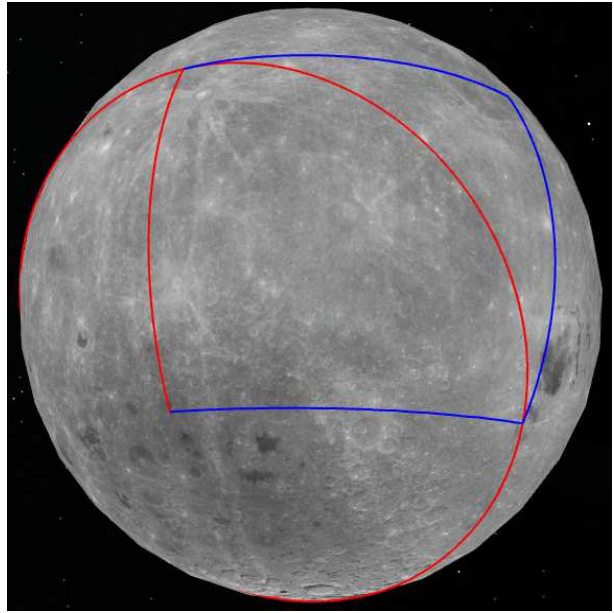}
\caption{Circle (red line) and square (blue line) inscribed in it considered
for the measurement of the number craters. The image was obtained from
Google Moon (see \protect\cite{google}).}
\label{moonAA}
\end{figure}

As it was said in last section, there is an interesting example in nature
where random points are inscribed in a spherical geometry: the Moon or any
planet with a large numbers of craters in its surface. There are zones in
the Moon where the craters are distributed almost randomly as it can be seen
in figure \ref{moonAA}, where it is shown the Moon with selected spherical
square obtained from Google Moon \cite{google} and the craters are marked
and can be seen in the Supplementary Material (\cite{sup}). The software
used \cite{google} allows to draw great circles, circles and spherical
squares with the respective lengths and areas. The zone in the Moon was
chosen due to the fact it is the largest spherical square located
symmetrically with respect a great circle of the Moon with the largest
number of almost random craters. 
Then, it is possible to implement the same procedure explained in the last
section by counting craters inside a spherical square and in a quarter of
circle inscribed in the surface of the Moon. The ratio obtained $\alpha
=4N_{C}/N$ depends on the radius of the quarter of circle $r$ chosen and the
Moon's radius $R_{M}$. This means that we can obtain an estimated value of
the Moon's radius by simply counting craters in a square over the surface.
This could sound peculiar but is a natural consequence of the Monte-Carlo
method in other geometries besides Euclidean geometry.

In order to do this, Google Moon \cite{google} was used, where a detailed
image is available and where the craters can be pointed by a mark. The
suplemmentary material (\cite{sup}) contains the circle and the spherical
square chosen and the marks of each crater. The yellow marks are the craters
inside the circle and the green marks are the craters outside the circle and
inside the square.

In order to apply the method explained above, a square of side $r=1815$ km
was considered as it can be seen in figure \ref{moonBBB}. The points where
craters are found are marked (see supplementary material \cite{sup}). By
counting the number of craters $N_{C}$ inside the quarter of circle
inscribed in the spherical square and the total number of craters $N$ inside
the square, the factor $4N_{C}/N$ can be computed and the value obtained is%
\begin{equation}
\alpha =\frac{4N_{C}}{N}=4\times \frac{1145}{1752}=2.616  \label{m0}
\end{equation}%
where the total number of random craters is $N=1752$.

By using eq.(\ref{3.2}) or by computing numerically the inverse function $%
\alpha $, the Moon's radius obtained is $R_{M}=1820$ km. By using the Taylor
expansion up to second order of eq.(\ref{3.2}), that is $\alpha \sim \pi -%
\frac{2}{3}\pi (\frac{r}{2R})^{2}$, the value of the Moon's radius reads $%
R_{M}=1811$ km.

\begin{figure}[tbp]
\centering\includegraphics[width=80mm,height=70mm]{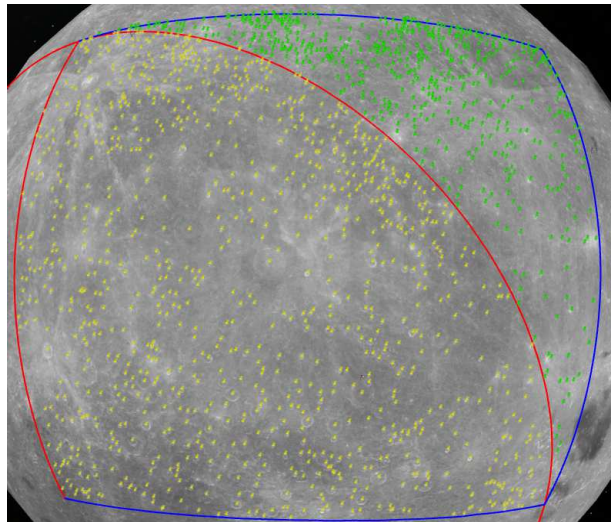}
\caption{Craters found in the circle and square over the Moon (image
obtained from Google Moon). }
\label{moonBBB}
\end{figure}

\begin{figure}[tbp]
\centering\includegraphics[width=100mm,height=70mm]{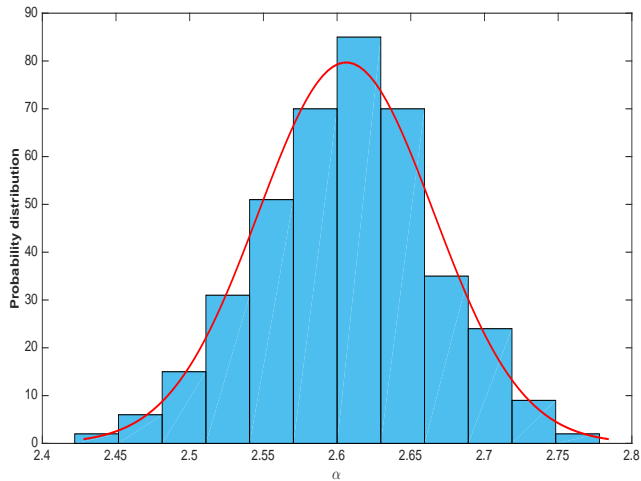}
\caption{Distribution function of the possible results of $\protect\alpha $
in the numerical simulation where $N=1752$ random points have been
considered, which is the number of random craters considered in the
experimental procedure. }
\label{gauss}
\end{figure}

Due to the fact that we have used $N=1752$ random points ($1752$ random
craters) the Moon's radius value obtained has an error. The same behavior is
obtained for the Monte-Carlo method in flat geometry to obtain $\pi $, as it
was shown in the last section, where the confidence bounds scales as $%
N^{-1/2}$. In order to obtain the associated error of $R_{M}$ that we call~$%
\Delta R_{M}$ in terms of the number of craters considered, we have
implemented numerically the Monte-Carlo method in spherical geometry with
the same parameters (inscribed radius circle $r=1815$ km) and we have
performed several calculations of $\frac{4N_{C}}{N_{S}}$ considering $N=1752$
random points (for the numerical implementation see Supplementary Material).
The dispersion of the results can be seen in figure \ref{gauss}. 
As it happens with the Monte-Carlo method to obtain $\pi $, the dispersion
of the results depends only on the number of random points. The same is
expected in spherical geometry, the dispersion of the results are
independent of $r$ and $R_{M}$ and only depends on the number of craters
considered for the calculation. Nevertheless, constant dispersion in $\alpha
=\frac{4N_{C}}{N_{S}}$ does not imply constant dispersion in $\frac{r}{2R_{M}%
}$. We might note this by computing the differential $d\alpha $ as $d\alpha =%
\frac{d\alpha }{d(\frac{r}{2R_{M}})}d(\frac{r}{2R_{M}})$ which implies that
for constant $d\alpha $ we have that $d(\frac{r}{2R_{M}})$ has a dependence
on $\frac{r}{2R_{M}}$. Considering the differential $d\alpha $ as the error
associated to $\alpha =\frac{4N_{C}}{N_{S}}$ written as $\Delta \alpha $ and
the differential $dR_{M}$ as the error in $R_{M}$ written as $\Delta R_{M}$,
a straighforward calculation gives the error associated to the Moon's radius%
\begin{equation}
\Delta R_{M}=\frac{1}{\pi }\left\vert \frac{1}{r}\sin (\frac{r}{R_{M}})-%
\frac{1}{R_{M}}\cos (\frac{r}{R_{M}})\right\vert ^{-1}\Delta \alpha
\label{error}
\end{equation}%
The error $\Delta \alpha $ is obtained from the tails of the dispersion of
figure \ref{gauss} when the Monte-Carlo method is implemented numerically.
Using that $\Delta \alpha =\sigma =0.045$ and using $r=1815$ km in last
equation the final result for the Moon radius is%
\begin{equation}
R_{M}=1820\pm 146\text{ km}  \label{res}
\end{equation}%
The obtained value for the Moon radius is accurate and the error contains
the real value of the Moon radius.

Should be stressed that a better approximation can be obtained for the mean
Moon radius $R_{M}$ by considering other spherical squares located
symmetrically with respect the great circle of the Moon where the craters
are randomly distributed. A simple inspection using Google moon software
shows that the largest zone is the one considered in this work.

\section{Conclusions}

In this work the Moon radius can be obtained experimentally by counting
craters in an spherical square and a quarter of circle inscribed in it. This
procedure is the generalization of the Monte-Carlo method to obtain $\pi $
in Euclidean geometry to spherical geometry. By deviating from flat
geometry, the ratio of random points in a square and a circle inscribed in
it gives a deviation from $\pi $. This deviation can be related to the
radius of inscribed circle and the radius of the sphere. In particular, the
Moon contain zones with random craters that can be used as random points
over a surface sphere. By applying the Monte-Carlo method, that is, by
counting craters inside the square and circle defined in the surface of the
Moon, a deviation from $\pi $ is obtained and this result is used to compute
the Moon radius. The obtained value is $R_{M}=1820\pm 146\text{ km}$ and the
real value of the Moon radius is inside the error. Although the method
introduced in this work is not very precise, shows how the randomness can be
useful to obtain information about the underlying space in which this random
phenomena occurs. In turn, this method can give better approximations by
considering several squares and circles inscribed in the Moon.

\section{Acknowledgment}

This paper was partially supported by grants of CONICET (Argentina National
Research Council) and Universidad Nacional del Sur (UNS) and by ANPCyT
through PICT 1770, and PIP-CONICET Nos. 114-200901-00272 and
114-200901-00068 research grants, as well as by SGCyT-UNS., J. S. A. is a
member of CONICET.

\section{Appendix A:\ Sides and area of the spherical square}

The line element in spherical coordinates with a fixed radius is $d\mathbf{l=%
}Rd\phi \widehat{e}_{\phi }\mathbf{+}R\sin \phi d\lambda \widehat{e}%
_{\lambda }$. Considering that two of the four sides of the spherical square
can be obtained with constant $\lambda $ and the two remaining sides with
constant $\phi $, we obtain by integration of $d\mathbf{l}$ between $-\phi
_{0}$ and $\phi _{0}$ in $\phi $ and between $0$ and $\lambda _{0}$ in $%
\lambda $, and making both results identical (see figure \ref{sphere}) 
\begin{equation}
r=2R\phi _{0}\text{ \ \ \ \ \ \ \ \ }r=R\lambda _{0}\cos \phi _{0}
\label{sq8}
\end{equation}%
where $r$ is the length of the side of the spherical square and
simultaneously is the radius of the circle inscribed in the square. Making
both results identical we obtain that%
\begin{equation}
\lambda _{0}=2\phi _{0}/\cos \phi _{0}  \label{s19}
\end{equation}%
This last result is used in section I.

In order to compute the area of the spherical square centered in a great
circle of radius $r$ inscribed in a sphere of radius $R$, the surface
element must be used $dA=R^{2}\sin \theta d\theta d\varphi $ where $\theta $
is the polar angle and $\varphi $ the azimuthal angle. By using the
longitude $\lambda =\varphi $ and latitude $\phi =\frac{\pi }{2}-\theta $,
the area of the square reads%
\begin{equation}
A_{S}=R^{2}\int_{\frac{\pi }{2}-\phi _{0}}^{\frac{\pi }{2}+\phi _{0}}\sin
\theta d\theta \int_{0}^{\lambda _{0}}d\varphi =2R^{2}\sin \phi _{0}\lambda
_{0}  \label{sq1}
\end{equation}%
In turn because the sides of the square must be identical to the radius~$r$
of the inscribed circle, then $r=R2\phi _{0}$ and $r=R\cos (\phi
_{0})\lambda _{0}$, which implies that $\lambda _{0}$ can be determined as $%
\lambda _{0}=2\phi _{0}/\cos (\phi _{0})$. Introducing this result in last
equation, the area of the spherical square reads%
\begin{equation}
A_{S}=4R^{2}\phi _{0}\tan (\phi _{0})  \label{sq2}
\end{equation}%
where $\phi _{0}=\frac{r}{2R}$. The area of the inscribed circle can be
computed by simply realizing that this circle is a spherical cap with an
angle $2\phi _{0}$ between the rays from the center of the sphere to the
pole and the edge of the disk forming the base of the cap. This area reads%
\begin{equation}
A_{C}=2\pi R^{2}\left[ 1-\cos (2\phi _{0})\right]  \label{sq3}
\end{equation}%
using that $r=R2\phi _{0}$ then $\phi _{0}=\frac{r}{2R}$ and the area of the
square and area of the circle can be written in terms of $r$ and $R$ as%
\begin{equation}
A_{S}=2rR\tan (\frac{r}{2R})  \label{sq4}
\end{equation}%
and%
\begin{equation}
A_{C}=2\pi R^{2}\left[ 1-\cos (\frac{r}{R})\right]  \label{sq5}
\end{equation}%
These results will be used in Section II. \

\bigskip
\bigskip
\textbf{Supplementary material: Numerical implementation of Monte-Carlo
method in spherical geometry}

As it was shown in Section II of the manuscript, the area of the square and
the circle in spherical geometry reads%
\begin{equation}
A_{S}=2rR\tan (\frac{r}{2R})  \label{a1}
\end{equation}%
and the area of a circle of radius $r$ inscribed in a sphere of radius $R$
reads%
\begin{equation}
A_{C}=\frac{\pi }{2}R^{2}\left[ 1-\cos (\frac{r}{R})\right]   \label{a2}
\end{equation}%
when $r/R\rightarrow 0$ both areas are $A_{S}=r^{2}$ and $A_{C}=\frac{\pi
r^{2}}{4}$ as it is expected and $\frac{4A_{C}}{A_{S}}=\pi $ holds. Then, by
using eq.(\ref{a2}) and eq.(\ref{a3}), the factor $4A_{C}/A_{S}$ reads%
\begin{equation}
\frac{4A_{C}}{A_{S}}=\frac{\pi R}{r}\sin (\frac{r}{R})  \label{a3}
\end{equation}%
and the result depends on the ratio $r/R$. In the limit $r/R\rightarrow 0$,
the function behaves as%
\begin{equation}
\frac{4A_{C}}{A_{S}}\sim \pi -\frac{2}{3}\pi (\frac{r}{2R})^{2}+\frac{2}{15}%
\pi (\frac{r}{2R})^{4}+O\left( (\frac{r}{2R})^{6}\right)   \label{a4}
\end{equation}%
The last equation shows the deviations from $\pi $ in the Monte-Carlo method
applied in spherical geometry and this deviation can be expanded in powers
of $\frac{r}{2R}$, which implies that by knowing how much we have deviate
from $\pi $ then we can deduce $R$ by an appropiate choice of $r$.

\begin{figure}[tbp]
\centering\includegraphics[width=90mm,height=80mm]{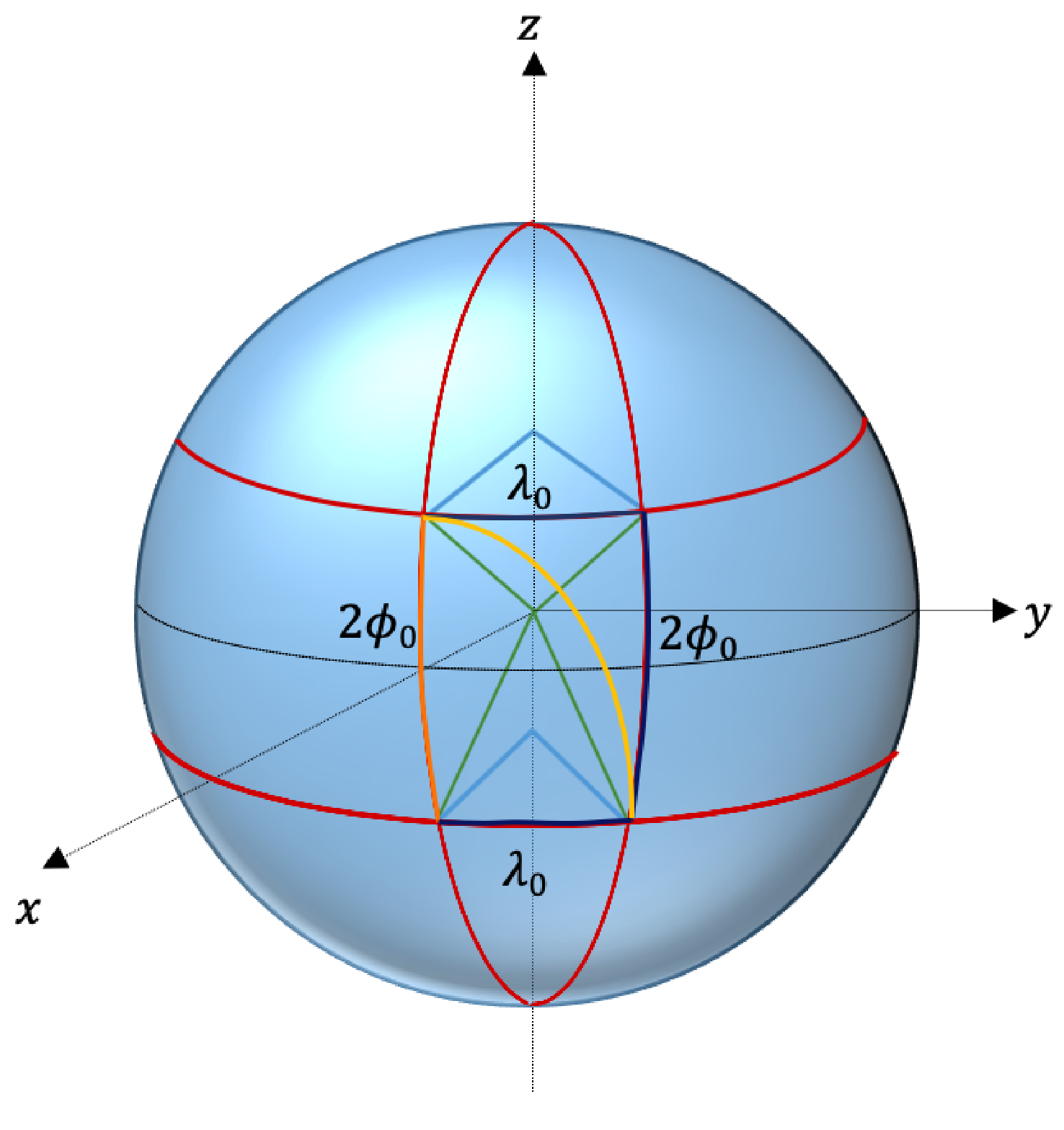}
\caption{Spherical square (blue line) located symmetrically with respect a
great circle of the sphere and the quarter of circle (yellow line) inscribed
in it. The sides of the spherical square are shown as $\protect\lambda _{0}$
and $2\protect\phi _{0}$. These two sides must be identical in order to have
a correct limit in flat geometry. }
\label{sphere2}
\end{figure}

In order to implement the Monte-Carlo method \ in spherical geometry
numerically, we can consider the square defined by the angles $\phi \in %
\left[ -\phi _{0},\phi _{0}\right] $, $\lambda \in \left[ 0,\lambda _{0}%
\right] $ in spherical coordinates, where $\phi _{0}$ is the maxium latitude
and $\lambda _{0}$ is the maximum longitude as it can be seen in the figure %
\ref{sphere2}. The angles $\phi _{0}$ and $\lambda _{0}$ are related by

\begin{equation}
\lambda _{0}=2\phi _{0}/\cos \phi _{0}  \label{3.3}
\end{equation}
which is obtained due to the fact that the sides of the spherical square are
identical, that is, $r=2R\phi _{0}$ and $r=R\lambda _{0}\cos \phi _{0}$. The
relation $r=2R\phi _{0}$ implies that $\phi _{0}$ can be written as 
\begin{equation}
\phi _{0}=\frac{r}{2R}  \label{3.4}
\end{equation}%
which implies that the latitude $\phi _{0}$ depends on the ratio $r/R$. In
turn, that the maximum value of the radius $r$ of the inner circle is $%
r_{\max }=\pi R$, which implies a maximum value of $\phi _{0}^{\max }=\pi /2$%
. This result is expected due to the fact that the radius $r$ of the
inscribed circle must be smaller than the perimeter of a great circle of the
sphere, which is identical to $\pi R$. When $r=\pi R$, the area of the
circle is half the area of the sphere. Moreover, the angle $\phi _{0}$ is
the angle formed by the rays from the center of the sphere to the edge of
the disk that the circle form in the sphere of radius $R$. Considering $N$
random points in the interval $\phi \in \left[ -\phi _{0},\phi _{0}\right] $
and $\lambda \in \left[ 0,2\phi _{0}/\cos \phi _{0}\right] $, deviations
from $\pi $ can be obtained when the ratio of the number of points that fall
inside the circle to the total number of points is computed and this
deviation depends exclusively on the ratio $r/R$. By knowing $r$ and $%
4N_{C}/N$, the value of $R$ can be obtained. Should be stressed that in
order to pick a random point on any surface of a unit sphere, is not correct
to consider the spherical coordinates $\theta $ and $\phi $ from uniform
distributions $\theta \in \left[ 0,2\pi \right] $ y $\phi \in \left[ 0,\pi %
\right] $, since the area element $dA=R^{2}\sin \phi d\theta d\phi $ is a
function of $\phi $ and then the points picked are clustered in the poles.
The correct choice is by introducing the distribution $\frac{1}{2}\sin
\varphi $ for the azimuth.

By writing eq.(\ref{3.1}) using eq.(\ref{3.4}) we obtain that the function $%
\alpha (\phi )$ reads%
\begin{equation}
\alpha (\phi )=\frac{\pi }{2\phi }\sin (2\phi )  \label{m1}
\end{equation}%
where $\phi =\frac{r}{2R}$ and $\alpha =$ $\frac{4A_{C}}{A_{S}}$ that in the
numerical implementation becomes $\frac{4N_{C}}{N_{S}}$ where $N_{C}$ is the
number of random points in the quarter of circle and $N_{S}$ is the number
of random points inside the square, which is identical to the total number
of random points $N$.

\begin{figure}[tbp]
\centering\includegraphics[width=120mm,height=80mm]{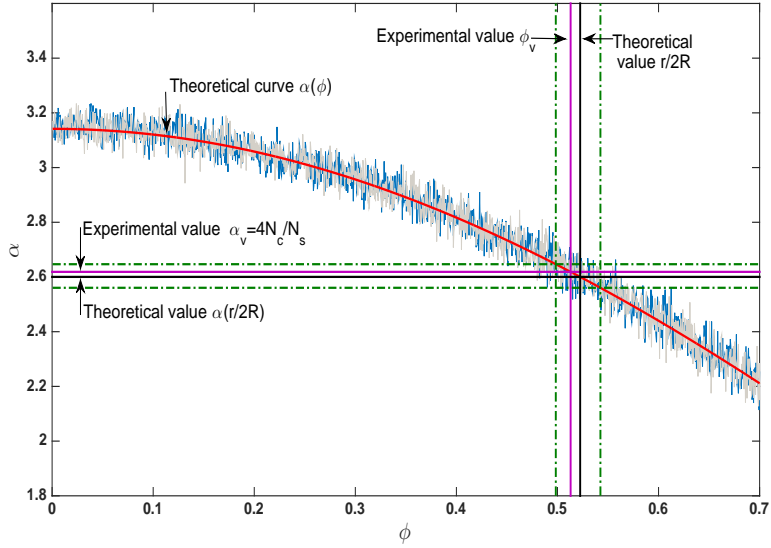}
\caption{Theoretical and numerical simulation of the function $\protect%
\alpha (\protect\phi )$ that accounts for the deviation of $\protect\pi $
when the geometry is curved. The horizontal axis is the angle $\protect\phi =%
\frac{r}{2R}$, where $r$ is the radius of the inscribed circle and $R$ is
the sphere radius. }
\label{alfafiplot}
\end{figure}

In figure \ref{alfafiplot}, the function $\alpha (\phi )$ is shown in red
and different numerical calculations show the dispersion of the result
around the theoretical curve, where we have considered $N=1752$ random
points over the spherical square inscribed in the sphere. The dispersion of
the results are independent of $\phi $ and only depends on the number of
random points $N$, as it happens with the Euclidean Monte-Carlo method to
measure $\pi $. Nevertheless, constant dispersion in $\alpha $ does not
implies constant dispersion in $\phi $. We might note this by considering
that $d\alpha $ is related to $d\phi $ as $d\alpha =\frac{d\alpha }{d\phi }%
d\phi $ which implies that for constant $d\alpha $ we have that $d\phi $ has
a dependence on $\phi $. This implies that once we obtain $d\phi $ in terms
of $\phi $, we can obtain the error in $R_{M}$ by $\left\vert d\phi
\right\vert =\frac{r}{2R_{M}^{2}}dR_{M}$. 
In the same figure \ref{alfafiplot}, the vertical black line indicates the
theoretical value is $\phi _{T}=\frac{r}{2R}=0.5224$, where $r=1815$ km and $%
R=1737$ km is the actual Moon's radius. The horizontal black line indicates
the value $\alpha (\phi _{T})$ and the dashed green lines indicates $\alpha
(\phi _{T})\pm \sigma $, where $\sigma $ is obtained from a Gauss
distribution of the possible $\alpha $ values when $\phi _{T}=0.5224$
considering $N=1752$ random points.

The tails $\alpha (\phi _{T})\pm \sigma $ are shown in green dashed lines in
figure \ref{alfafiplot}. In turn, vertical dashed green lines are shown
around $\phi _{V}\,\ $which indicates the dispersion expected in the $\phi $
axis. Finally, the violet horizontal line indicates the obtained
experimental value $\alpha _{V}=\frac{4N_{C}}{N_{S}}=2.61$ in the manuscript
by counting craters in the surface of the Moon, which gives $\alpha _{V}=%
\frac{4N_{C}}{N_{S}}=4\times \frac{1145}{1752}=2.616$ and where we have $%
N=1752$ random craters and the radius of the circle inscribed in the Moon's
surface is $r=1815$ km. The vertical violet line is the experimental value
by inverting the function $\alpha (\phi )$, that is $\phi _{V}=\alpha
^{-1}(\alpha _{V})$. From the figure \ref{alfafiplot} it can be seen that
the experimental value obtained for $\phi _{V}$ is inside the vertical
dashed lines. By inverting eq.(\ref{m1}), the value obtained for $\phi _{V}$
is $\phi _{V}=0.5178$ and the Moon radius reads $R_{M}=1820$ km. Finally, in
order to obtain the error in the $R$, we can consider that the possible
values of $\alpha $ implies possible values of $\phi $ implicitly through
eq. (\ref{m1}) for a fixed number of random points $N$. We can call the
error of the variable $\alpha $ as $\Delta \alpha $ and the error of $\phi $
as $\Delta \phi $ and both are related through eq.(\ref{m1}) as%
\begin{equation}
\Delta \alpha =\frac{d\alpha }{d\phi }\mid _{\phi =\phi _{T}}\Delta \phi
\label{err1}
\end{equation}%
then%
\begin{equation}
\Delta \phi =(\frac{d\alpha }{d\phi }\mid _{\phi =\phi _{T}})^{-1}\Delta
\alpha  \label{err2}
\end{equation}%
and by using that $\phi _{T}=\frac{r}{2R_{M}}$%
\begin{equation}
\Delta R_{M}=\left\vert \frac{r}{2\phi _{T}^{2}}(\frac{d\alpha }{d\phi }\mid
_{\phi =\phi _{T}})^{-1}\right\vert \Delta \alpha  \label{err3}
\end{equation}%
The error in $\Delta \alpha $ obtained with $N=1752$ random points used is $%
\Delta \alpha =\sigma =0.045$.In order to compute the error associated to
this value, we should translate the dispersion in $\alpha $ given be $\sigma 
$ to the dispersion in $\phi $. Using that $\Delta \alpha =\sigma =0.045$
and that $d\alpha =\frac{d\alpha }{d\phi }d\phi $ and $\left\vert d\phi
\right\vert =\frac{r}{2R_{M}^{2}}dR_{M}$ we obtain the error in $R_{M}~$as%
\begin{equation}
\Delta R_{M}=\left\vert \frac{r}{2\phi _{T}^{2}}(\frac{d\alpha }{d\phi }\mid
_{\phi =\phi _{T}})^{-1}\right\vert \Delta \alpha  \label{er3.0}
\end{equation}%
By using that $\frac{d\alpha }{d\phi }=\frac{\pi }{\phi }\left[ \cos (2\phi
)-\frac{1}{\phi }\sin (2\phi )\right] $ and $r=1815$ km, then the final
result for the Moon radius is%
\begin{equation}
R_{M}=1820\pm 146\text{ km}  \label{er4}
\end{equation}

\bigskip 

\end{document}